\def\ii#1\ff{\textul{#1}}	

\def\diff{\mathrm{d}}

\def\A{\mathrm{A}}
\def\B{\mathrm{B}}
\def\C{\mathrm{C}}
\def\D{\mathrm{D}}

\def\vecp{\mathbf{p}}

\def\Ntest{N_{\textrm{test}}}

\def\bbsty#1#2#3{#1 (#3) #2}	
\def\textsty#1{#1}	

\documentclass[final,5p,times,twocolumn, psfig,amsmath,amssymb]{elsarticle_modif}
\usepackage{graphicx}\usepackage{epsfig}
\usepackage{amsmath}\usepackage{amssymb}\usepackage{amsfonts}
\usepackage{bm}
\usepackage{exscale}
\usepackage{dcolumn}    
\usepackage{color}
\usepackage{soul}
\usepackage{ulem}
\usepackage[numbers]{natbib}
\usepackage{fancyhdr}
\usepackage{hyperref}
\begin{document}
%
%
\title{\flushleft \scalebox{1.55}[1.55]{Nuclear jets in heavy-ion collisions recall a stream of sand}}
%
%
\author[1]{\large{P. Napolitani}}
\author[2]{\large{M. Colonna}}
%
%
\address[1]{IPN, CNRS/IN2P3, Universit\'e Paris-Sud 11, Universit\'e Paris-Saclay, 91406 Orsay Cedex, France}
\address[2]{INFN-LNS, Laboratori Nazionali del Sud, 95123 Catania, Italy}
%
%
\begin{abstract}
%
%
	Head-on collisions between nuclei of different size at Fermi energies may give rise to extremely deformed dynamical regimes and patterns.
	Those latter, may suddenly turn into a stream of nuclear clusters, resembling collimated jets.
	Because the underlying instabilities are inadequately described by usual modelling approaches based on equilibrium approximations, this mechanism resulted rather unnoticed, even though it should be frequently registered in experiments.
	We employ the Boltzmann-Langevin equation to specifically address out-of-equilibrium conditions and handle dynamical fluctuations.
	An interesting interplay between surface and volume instabilities is discussed for the first time.
	Stable and rather regular patterns of streaming clusters arise from these conditions.
	Counterintuitively, we find that these clustered structures are not triggered by cohesive forces and they recall the granular flow of a stream of dry sand.

\end{abstract} 
%
%

\maketitle

%
%
%
%
%

\section{Introduction}

	From microphysics to cosmological scale, jet regimes are frequent.
	They are collimated streams of matter which eventually clusterise into packets, yielding a variety of nonlinear behaviours~\cite{Eggers2008,Cross1993,Scott2007}.
	Since early nuclear-fission models~\cite{Vandenbosch1973,Griffin1976,Blocki1978}, the occurrence and rupture of deformed stretched structures in nuclear reactions, like neck configurations, suggested the analogy to viscous liquids subject to the Rayleigh instability~\cite{Rayleigh1882,Brosa1990,Gopan2014}, whereby a fluid thread breaks up into droplets.
	Such analogy became emblematic~\cite{Montoya1994,Toke1995,Lecolley1995} in explaining nuclear reactions from low-energy to Fermi energy (i.e. above about 30 MeV per nucleon).
	We explore the possibility that, beyond Fermi energy, columnar configurations could arise in heavy-ion collisions from different conditions than those producing a neck in dinuclear systems~\cite{DiToro2006,Baran2012}, and that they undergo rupture from mechanisms other than Rayleigh-type instabilities, i.e. from mechanisms that are almost unrelated to cohesive properties.
	In analogy with columnar streams of matter which are widely encountered in nature, from liquid to granular flows, we may generally refer to these configurations as nuclear jets.
	By employing a theoretical approach where we add collisional correlations and Langevin-type fluctuations to a mean-field description, we found that nuclear jets are frequent in collisions of nuclei of asymmetric size around Fermi energy.
	From the analysis of the type of instability which triggers rupture, 
we advance the conclusion that the clusterisation of the jet reflects an interesting interplay between volume and surface instabilities, with a leading role played by nucleon-nucleon (n-n) correlations and volume instabilities.
	Quite seemingly, analogue situations are experienced in granular jets of dry sand which, even in vacuum, can breakup into packets after some distance is travelled~\cite{Lohse2004,Lohse2004b,Royer2005,Royer2008}.
	In absence of surface tension, both mechanisms arise in fact from correlations beyond one-body dynamics~\cite{Royer2009}.
	More specifically, we find that clusterisation along the jet selects small sizes, favouring light nuclear clusters.
	This may lead to a new production mechanism of exotic clusters.
%


\section{Modelling nuclear jets}

	Widely used in many branches of physics, microscopic theories of transport phenomena are applied to heavy-ion collisions in a broad range of energies.
%
%
	For instance, nuclear transport dynamics can be put in the form of a kinetic equation where the propagation of the one-body density $f$ in phase-space depends upon the effective Hamiltonian $h[f]$ and the contribution of n-n collisional correlations.
	To describe nuclear jets we should explain how clusterisation progresses in colliding nuclei.
	Such a chaotic mechanism could be handled by stochastic transport theories~\cite{Lacroix2014} as those applied to diffusive processes, like the Brownian motion~\cite{Chomaz1994}.
	The additional degrees of freedom involved in turning the jet into a stream of clusters can be introduced in approximate form by the action of a fluctuating seed on the one-body density, so that a single mean-field path $f$ taken at a given time, splits into a subensemble of new trajectories $f^{(n)}$ which propagate at successive times.
	Such scheme is represented by the Boltzmann-Langevin equation~\cite{Ayik1988,Reinhard1992}
\begin{equation}
	\frac{\partial f^{(n)}}{\partial t} = \{h[f^{(n)}] , f^{(n)}\} 
		+ \bar{I}_{\textrm{coll}}[f^{(n)}] + \delta I_{\textrm{coll}}[f^{(n)}]   
	\;,
\label{eq:BL}
\end{equation}
where the Langevin term $\delta I_{\textrm{coll}}[f^{(n)}]$ acts as a fluctuating contribution around the average collision term $\bar{I}_{\textrm{coll}}[f^{(n)}]$, in the spirit of the Brownian motion.
The Langevin term leads to a diffusion coefficient $\mathcal{D}_{\textrm{coll}}$; it can produce bifurcations, and revives fluctuations intermittently in time.

	Clusterisation is a general catastrophic process characterising Fermi liquids~\cite{Pines1966}, stemming from conditions of instability and fluctuations.
	As a response, a combination of several fluctuation modes of large-amplitude are induced, where neutrons and protons may oscillate in phase or out of phase.
	In this process, the most amplified wavelengths are reflected into density ripples and, finally, into fragment formation.
	Their sizes has been found to characterise small atomic nuclei~\cite{Ayik1995,Chomaz2004,Borderie2018}.
	From the out-of-phase oscillations of neutrons and protons another process, isospin distillation, arises in relation to the nuclear symmetry energy, 
which leads to more symmetric fragments, thus affecting the final isotopic distributions of reaction products~\cite{Baran2012,Colonna2008}.
	Clusterisation can be efficiently handled by Eq.~(\ref{eq:BL}).
	We introduced the Boltzmann Langevin One Body (BLOB) approach~\cite{Napolitani2013,Napolitani2017} as a corresponding numerical realisation in three dimensions.
	Both isoscalar and isovector mechanisms (i.e. fragment formation and distillation) are satisfactorily handled with BLOB in nuclear matter~\cite{Napolitani2017}.
	We employ this approach to gain insight into the jet mechanism, and the underlying instability.
For comparison, we also employ an approximate version of stochastic mean field~\cite{Colonna1998}, where we suppress both n-n collisions and related fluctuations in the residual contribution of Eq.~(\ref{eq:BL}) and replace them by simple coarse-grained effects related to the mean field implementation $\bar{I}_{\textrm{coll}} + \delta I_{\textrm{coll}} \rightarrow \delta I_{\textrm{noise}}$.
	This collisionless approach is still able to activate breakup mechanisms but, with respect to the BLOB approach, it leads to a drastically reduced effective diffusion coefficient $\mathcal{D}_{\textrm{noise}} \ll \mathcal{D}_{\textrm{coll}}$. 
	See \textsection~\ref{supplemental} for simulation parameters.

\section{Results}

%
%
\begin{figure}[b!]
\begin{center}
	\includegraphics[angle=0, width=.99\columnwidth]{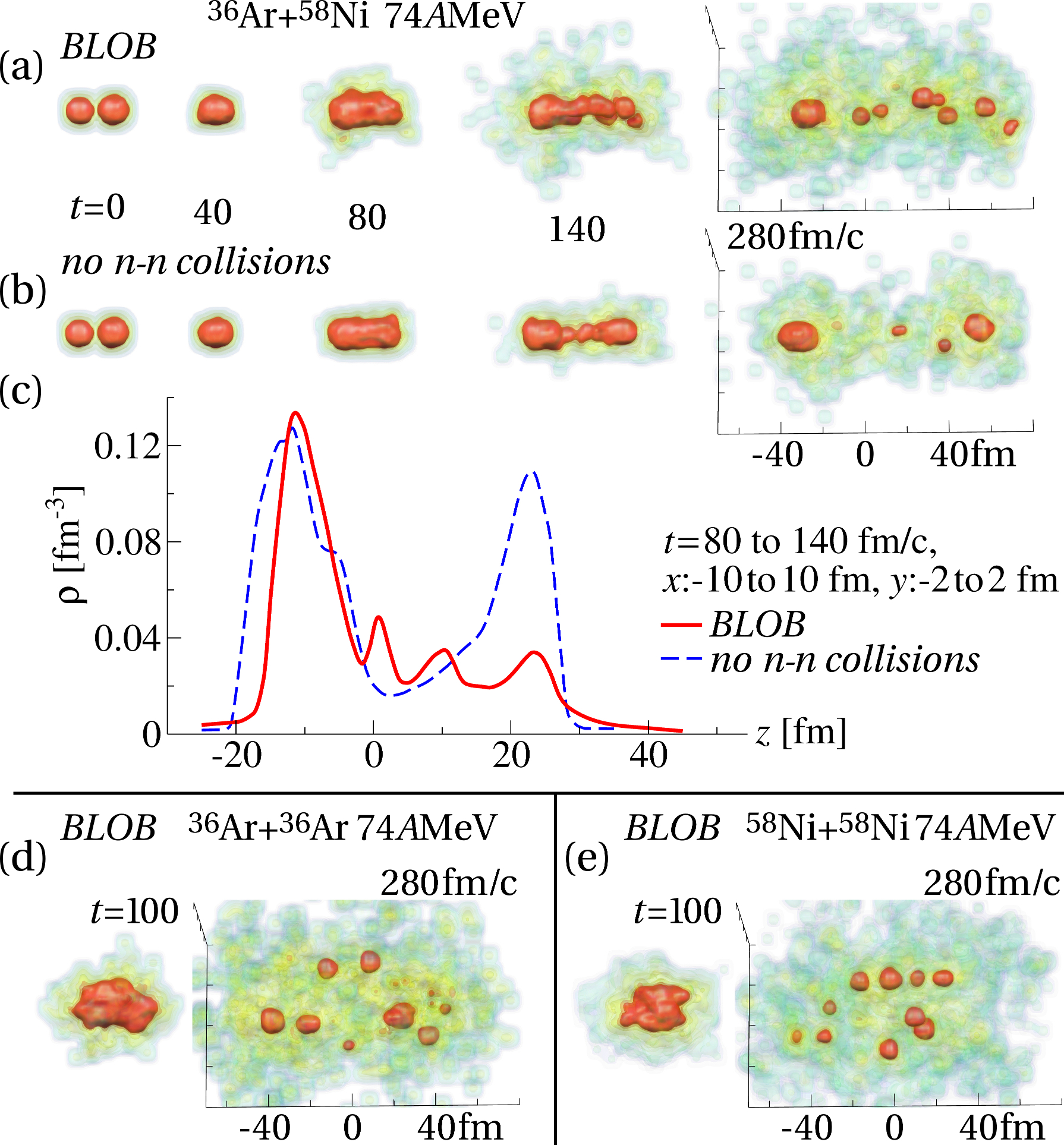}
\end{center}
\caption
{
	Simulation of $^{36}$Ar$+^{58}$Ni at 74 $A$ MeV with BLOB {(a)} and the collisionless approach {(b)}, for one single event.
	{(c)}, Corresponding density profile evaluated 
around the collision axis $z$ (cell size: 20fm in impact-parameter direction $x$ and 4fm in out-of-reaction-plane direction $y$) and averaged in a time span from 80 to 140fm/c.
	{(d,e)}, BLOB simulations of 
	$^{36}$Ar$+^{36}$Ar and $^{58}$Ni$+^{58}$Ni at 74 $A$ MeV.
} \vspace{-1ex}
\label{fig1}
\end{figure}
%
%
	
%
%
\begin{figure}[t!]
\begin{center}
	\includegraphics[angle=0, width=.99\columnwidth]{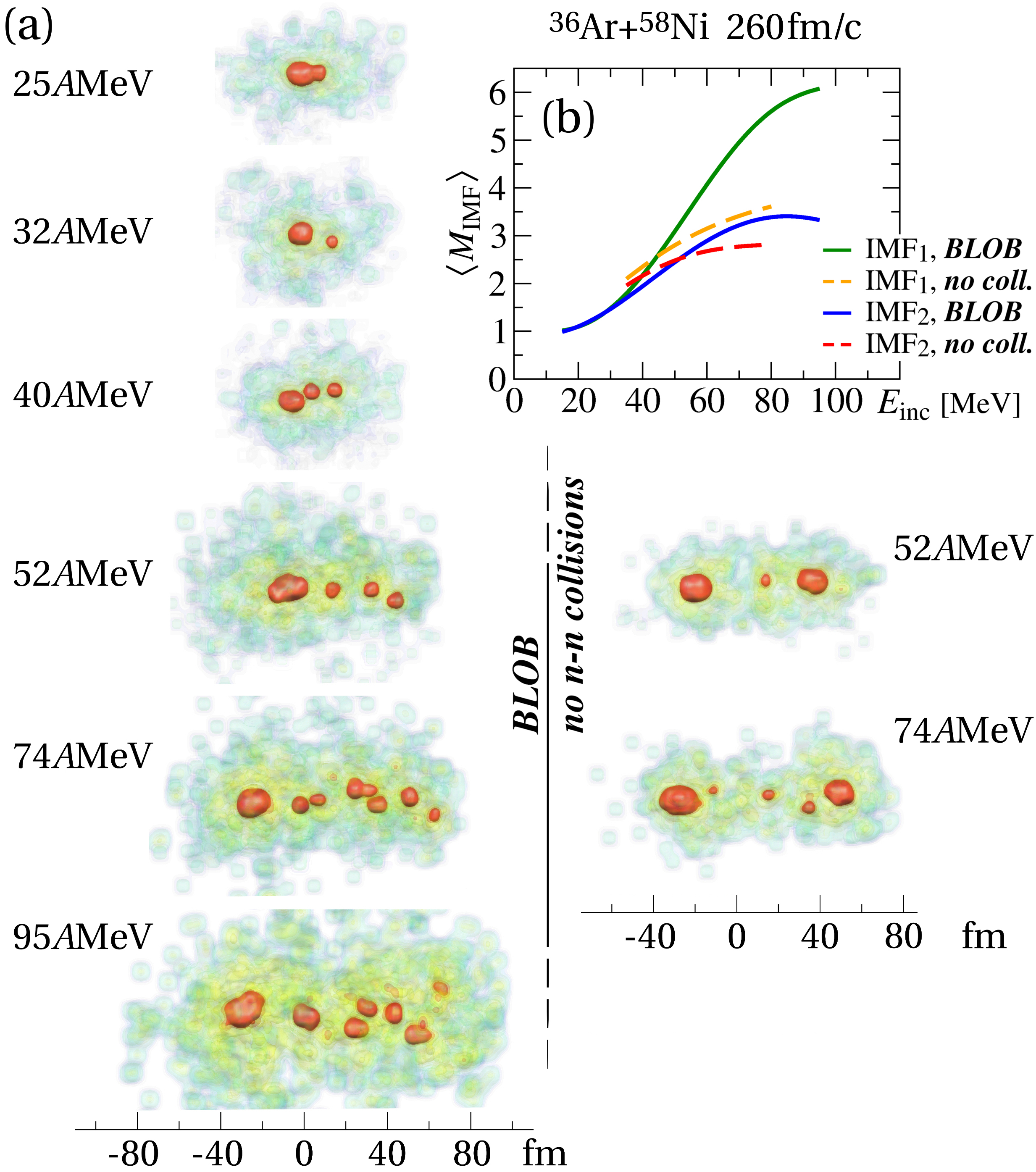}
\end{center}
\caption
{
	{(a)} Single events simulated with BLOB and the collisionless approach for $^{36}$Ar$+^{58}$Ni at 260fm/c as a function of bombarding energy; the event at 74~$A$MeV was presented in its full time evolution in Figs.~\ref{fig1}(a) and \ref{fig1}(b).
%
%
	{(b)} average multiplicity of fragments.  
$n$ and $p$ are excluded in the selection IMF$_1$. $n$,$p$,$d$,$t$,$^3$He,$\alpha$ are excluded in the selection IMF$_2$.
} \vspace{-1ex}
\label{fig2}
\end{figure}

	When atomic nuclei of different size are involved in head-on collisions above Fermi energy, 
the heaviest nucleus is heated up by the collision without suffering drastic modifications and it produces the heaviest residue in the exit channel: we name it $A_1$.
	Contrarily, the lighter partner tends to disintegrate into a jet of fast-streaming low-density matter, producing the second, third, etc. largest fragments, named $A_2$, $A_3$, etc., respectively.
	This is the pattern which has been reported in experiments, e.g. in ref.~\cite{Lautesse2006}, and which is also observed in BLOB simulations.
	As a detailed example, we analyse thereafter a typical system where the jet process should occur.
	A calculation illustrated in Fig.~\ref{fig1} tracks the evolution of the density distribution in configuration space in the asymmetric system $^{36}$Ar$+^{58}$Ni at 74 $A$ MeV (head-on collisions) and in the corresponding symmetric systems $^{36}$Ar$+^{36}$Ar and $^{58}$Ni$+^{58}$Ni for single stochastic events, chosen among the most probable ones.
	While the symmetric channels in Fig.~\ref{fig1}(d) and \ref{fig1}(e) manifest significant radial expansion, which is the well known signature of multifragmentation and vaporisation mechanisms~\cite{EPJAtopicalWCI2006,Colonna2017}, the asymmetric channels in Fig.\ref{fig1}(a) exhibit a columnar jet formation in the forward sector with respect to $A_1$.
	Since early times, the jet experiences a density drop along the longitudinal axis within $1/3$ to $1/8$ of the nuclear saturation density $\rho_{\textrm{sat}}$, as shown in Fig.~\ref{fig1}(c).
	The collisionless approach, Fig.~\ref{fig1}(b) leads to 
a neck-like pattern instead, where one or few small fragments are situated at intermediate rapidity with respect to two larger fragments 
%
%
%
\begin{figure}[b!]
\begin{center}
	\includegraphics[angle=0, width=.99\columnwidth]{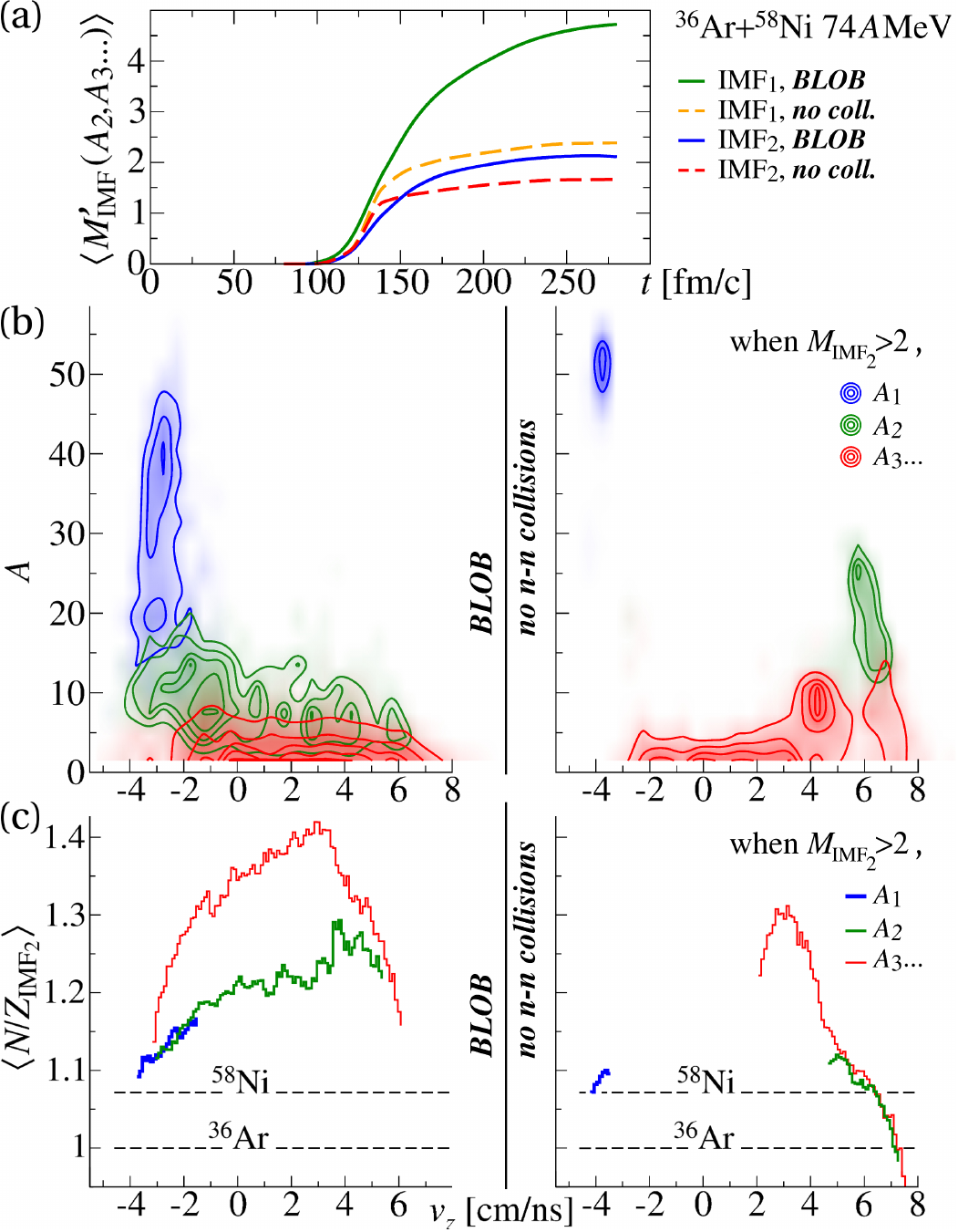}
\end{center}
\caption
{
	In $^{36}$Ar$+^{58}$Ni at 74~$A$MeV.
	{(a)} Time evolution of the average multiplicity of fragments of smaller size than $A_1$. 
	selections IMF$_1$ and IMF$_2$ as in Fig.~\ref{fig2}(b).
	{(b)} Correlation between fragment mass and longitudinal velocity $v_z$ at $t_{\textrm{stop}}$, for the distribution of $A_1$, $A_2$, and the remaining lighter fragments ($A_3...$) in events where at least three fragments belonging to IMF$_2$ are observed.
	{(c)} Isotopic content 
averaged over IMF$_2$ as a function of $v_z$ at $t_{\textrm{stop}}$. Same selections as in (b).
} \vspace{-1ex}
\label{fig3}
\end{figure}

	A survey of the most frequent configurations of forward fragment emission in $^{36}$Ar$+^{58}$Ni is proposed in Fig.~\ref{fig2} as a function of the incident energy.
	In Fig.~\ref{fig2}(a), jets from BLOB calculations
	arise above 40$A$MeV and dominate the exit channel from 52 to 74$A$MeV, while at lower energy binary and neck-like mechanisms gain larger share.
	At larger bombarding energy jets expand in elongation and width, displaying a more turbulent pattern.
	Fig.~\ref{fig2}(b) tracks the multiplicity of intermediate-mass fragments (IMF) according to two selections
labelled IMF$_1$ when accounting for any fragment other than free nucleons and protons, and IMF$_2$ when deuterons, tritons, $3$He and $\alpha$ particles are additionally excluded.
	At variance with the collisionless approach, the BLOB simulation yields different populations IMF$_1$ and IMF$_2$.
	The larger growth rate and earlier saturation of the population IMF$_2$ with respect to IMF$_1$ as a function of the incident energy, indicates that the production of light charged particles (LCP) contributes largely to the IMF multiplicity, involving more and more mass from the target-like nucleus.

	There is actually a fundamental difference between neck fragments and IMF arising along a jet, as examined in Fig.~\ref{fig3} for the system at 74~$A$MeV.
	Relying either on BLOB or on the collisionless version, Fig.~\ref{fig3}(a) traces the time evolution of the average multiplicity of fragments where $A_1$ is excluded, selecting either the population IMF$_1$ or IMF$_2$.
	The initial rapid growth of multiplicity simply indicates when projectile and target are crossing each other.
	At later times, in the collisionless approach the further evolution of multiplicity is slow and reflects a chiefly binary split with the possible presence of one or more light fragments.
	On the other hand, the BLOB calculation suggests that the multiplicity continues to grow till around 280fm/c due to the production of LCP.
	From this evolution, we can associate the beginning of instability growth, discussed in the following, to the time interval ranging from 110 to 120 fm/c.
	The mass of the fragments $A_1$, $A_2$, $A_3$, etc. is analysed in Fig.~\ref{fig3}(b) as a function of the longitudinal velocity $v_z$ (along the reaction axis) for events where at least three fragments are produced before the average multiplicity evolution flattens.
	In the collisionless approach we found that $A_2$ is characterised by the largest forward velocity, followed by smaller fragments, recalling the physics of peripheral or semicentral collisions~\cite{lukasik1997,Colin2003}, where cohesive properties, related to nuclear surface tension, have usually been invoked to explain the formation of clusters.
	On the contrary, the BLOB calculation suggests a reverse hierarchy, where larger longitudinal velocities tend to be correlated to smaller sizes; the same picture can be drawn from the density profile in Fig.~\ref{fig1}(c), where a series of maxima correspond to inhomogeneities and nesting places for clusters along a jet.
	Such behaviour is typically related to volume instability, and not to cohesive properties.

	The drop to low density 
also triggers the occurrence of isospin effects~\cite{Chomaz2004,Baran2005,Colonna2006,Coupland2011,EPJAtopicalNSE2014}
both in the jet formation and in the neck mechanism. 
	The average isospin content $<\!N/Z\!>$ of $A_1$, $A_2$ and of the lighter fragments ($A_3$...) is tracked in Fig.~\ref{fig3}(c) for events where at least three fragments are formed.
	In the collisionless approach, lighter fragments ($A_3$...) result more neutron rich than the isospin content of the system, reflected by $A_1$ and $A_2$, as expected for a migration process towards the low-density neck~\cite{Lionti2005,Baran2005,Ditoro2006bis}.
	On the other hand, in the BLOB approach, the $A_2$ fragments are more neutron rich than the system and the other lighter fragments reach even higher values.
	This effect results from an intense prompt emission of protons and $\alpha$ particles, and isospin distillation~\cite{Colonna2008,Ducoin2007}: 
this is one more indication that the jet could involve volume instability.

\section{Discussion: surface instability or volume instability?}
%
%
\begin{figure}[b!]
\begin{center}
	\includegraphics[angle=0, width=1\columnwidth]{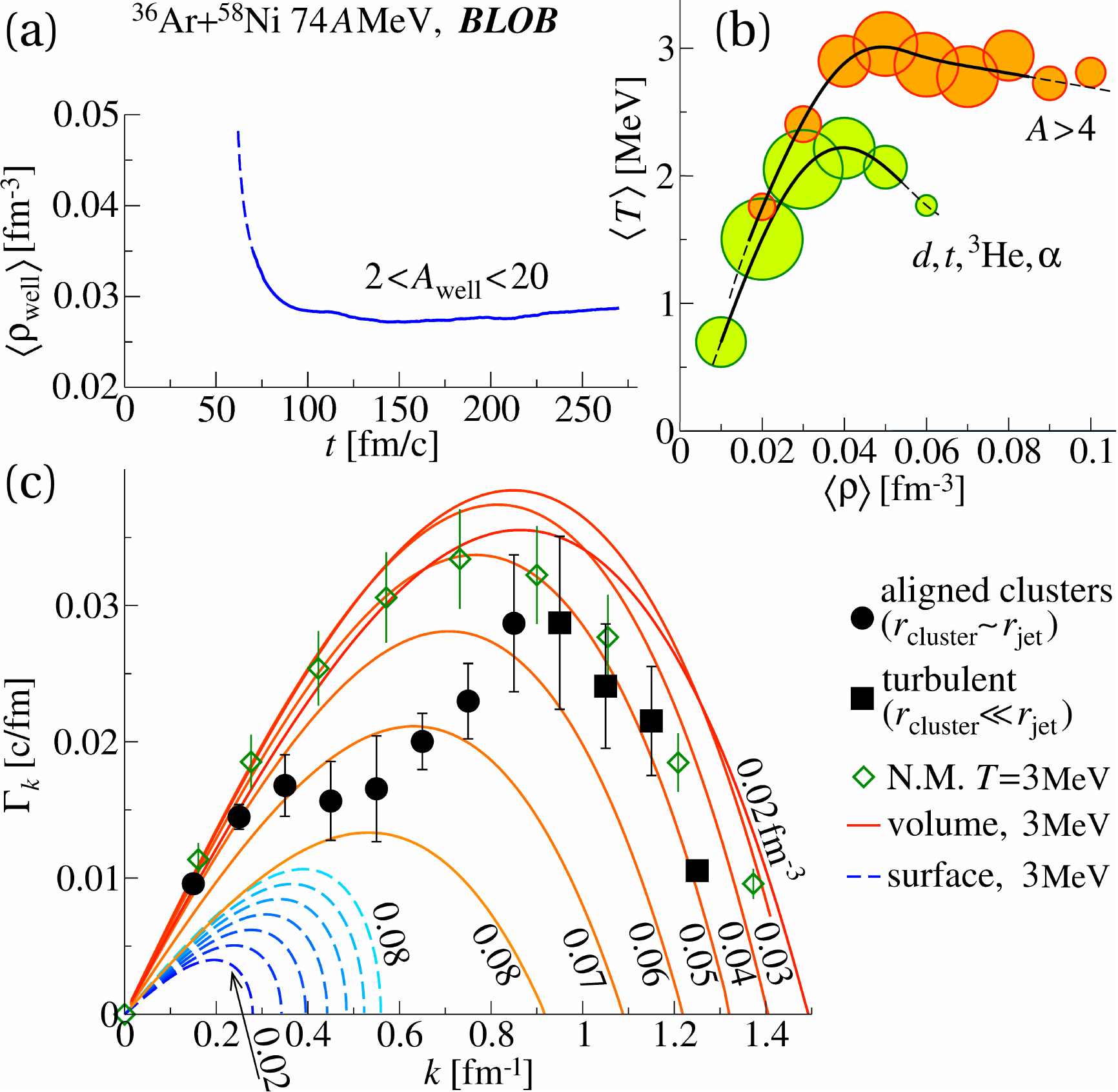}
\end{center}
\caption
{
	In $^{36}$Ar$+^{58}$Ni at 74~$A$MeV.
	{(a)} Time evolution of the density $\langle\rho_{\textrm{well}}\rangle$ of 
density ripples 
associated with 
nesting fragments of size $A_{\textrm{well}}$.
	{(b)} Correlation between temperature and density for 
light clusters ($d$, $t$, $^3$He, $\alpha$) and 
remaining IMFs
($A>4$) found in the jet ($A1$ is excluded) in a time range from 150 to 280 fm/c; symbol areas are normalised to total production yields.
	{(c)} Growth time as a function of the $k$ number for 
two jet splitting geometries (see text);
error bars are related to the initial time defining the onset of instabilities, 
ranging from 110 to 120 fm/c. 
	For comparison, analytic dispersion relations at $T=3$MeV and at different densities for 
volume (i.e. spinodal) and surface (i.e. Plateau-Rayleigh) instabilities are shown.
	BLOB calculations in nuclear matter (N.M.) at $\rho_{\textrm{sat}}/3$ and $T=3$MeV, from ref.~\cite{Napolitani2017}, are also reported.
} \vspace{-1ex}
\label{fig4}
\end{figure}
	From the above analysis, two types of instabilities emerge as possible contributions to clusterisation, according to whether they prevalently involve the surface or the volume of the jet.
	To explore their mutual role, fig.~\ref{fig4}(a) tracks the density averaged over emerging potential ripples $\langle\rho_{\textrm{well}}\rangle$ in the jet region as a function of time, calculated with BLOB for the system $^{36}$Ar$+^{58}$Ni at 74~$A$MeV.
	$\langle\rho_{\textrm{well}}\rangle$ suddenly drops to around a fourth of $\rho_{\textrm{sat}}$.
	When these ripples separate into fragments of mass $A$, their temperature $\langle T\rangle$ evolves with density $\langle\rho\rangle$ as shown in Fig.~\ref{fig4}(b).
	IMF with $A>4$ are formed at temperatures around 3 MeV in a wide density region, from $\rho_{\textrm{sat}}/2$ to $\rho_{\textrm{sat}}/4$, while lighter clusters ($d$, $t$, $^3$He, $\alpha$) are produced at smaller temperatures in density tails below $\rho_{\textrm{sat}}/5$.
	While the largely deformed columnar-like configuration may suggest a Plateau-Rayleigh surface instability, these density and temperature conditions are compatible with a volume instability of spinodal type.
	Fig.~\ref{fig4}(c) compares analytic expectations for both instabilities at different densities, and for a temperature of $3$ MeV, consistently with the production of IMF with $A>4$.
	In nuclear reactions, the Plateau-Rayleigh instability is commonly envisaged in nuclear fission or in the breakup of a neck.
	A schematic dispersion relation, see dashed lines in Fig.~\ref{fig4}(c), relates the growth rate for an unstable mode of wavenumber $k$ to the surface tension $\gamma$ and to the geometric properties of a columnar configuration of radius $r$ as~\cite{Brosa1990}
\begin{equation}
	(\Gamma_{k, \textrm{surf}})^2 = 
	\frac{\gamma}{\rho m r^3} \frac{I_1(kr)}{I_0(kr)} kr (1-k^2r^2)
	\;,\label{eq:Rayleigh}
\end{equation}
where $I_0$ and $I_1$ are modified Bessel functions and $m$ is the nucleon mass.
	The expression of the surface tension $\gamma$ should take into account the low local density $\rho$, as well as the isospin $\beta=(\rho_n-\rho_p)/\rho$ of the emerging fragments (Fig.~\ref{fig3}) and the finite temperature $T$ (see \textsection~\ref{supplemental}).
	The volume instability relies most efficiently to regions of the equation of state where nuclear incompressibility is negative (pressure reducing with increasing density), resulting in instabilities of spinodal type~\cite{Colonna1997,Chomaz2004}.
	It has been widely investigated in nuclear systems at low density which disintegrate in several similar-size fragments~\cite{Borderie2018}.
	In nuclear matter~\cite{Colonna1994_a,Colonna1994,Napolitani2017}, within the linear-response approximation, the dispersion relation, see solid lines in Fig.~\ref{fig4}(c), is
\begin{equation}
	1+\frac{1}{\tilde{F_0}(k,T)} =
		\frac{\Gamma_{k, \textrm{vol}}}{k v_{\textrm{F}}} 
		\textrm{arctan}\Bigg(
		\frac{k v_{\textrm{F}}}{\Gamma_{k, \textrm{vol}}} 
		\Bigg)
	\;,\label{eq:spinodal}
\end{equation}
where $v_{\textrm{F}}$ is the Fermi velocity and
$\tilde{F_0}(k,T) = (\mu(T)/\epsilon_{\textrm{F}}) F_0 g(k)$
is the effective Landau parameter including a dependence on temperature, through the chemical potential $\mu(T)$, the Fermi energy $\epsilon_{\textrm{F}}$, and the range of the nuclear interaction via the term $g(k)$; the range dependence (see \textsection~\ref{supplemental}) imposes an ultraviolet cutoff which bends the dispersion relation back to zero at small wavelengths (or large $k$).
	A BLOB calculation of the spinodal instability in nuclear matter at $\rho_{\textrm{sat}}/3$ and $3$ MeV is added for comparison (from ref.~\cite{Napolitani2017}).

	We compared the above analytic prescriptions to numerical results obtained with BLOB for the jet mechanism in $^{36}$Ar$+^{58}$Ni at 74~$A$MeV, as shown in Fig.~\ref{fig4}(c).
	To extract the dispersion relation, we applied a first-order analysis which, despite yielding possible underestimation of $\Gamma_{k}$, has the virtue of obtaining the dispersion relation directly from cluster correlations and the associated chronology in an open finite system.
	The breakup time of the jet $t_{\textrm{split}}$ is the average separation time of the clusters emerging from the jet calculated since inhomogeneities start to arise.
	The jet is approximated to a cylinder of length $l_{\textrm{jet}}$ equal in mass to the stream of clusters $A_i$ which travel in forward direction and which do not include the heavy residue $A_1$. 
The local density $\rho$ is averaged over the jet volume, and the radius $r$ is promptly obtained.
Further restrictions select only jets containing at least three clusters.
	In configurations where clusters are regularly aligned along the jet, we assume that the wavelength of the instability simply corresponds to the average spacing of two close clusters contained in the jet, see circles in Fig.~\ref{fig4}(c).
	On the other hand, when too many almost equal-size clusters fill the jet in a disordered pattern, in a sort of turbulent regime, the wavelengths do not include any stretching effect and they are therefore directly extracted from the average size of the emerging clusters,
see squares in Fig.~\ref{fig4}(c).
	We find that the dispersion relation obtained from the jet fragmentation is clearly outside the surface instability region, favouring volume instabilities.
	Even though it is rather similar to the BLOB result in nuclear matter at $\rho_{\textrm{sat}}/3$, some remarkable anomalies appear.
	The combination of many densities (with $\Gamma_{k, \textrm{vol}}$ decreasing for larger $\rho$) shifts the maximum towards smaller wavelengths (larger $k$).
	Moreover, a backbending appears for small $k$ modes, indicating 
the co-existence of volume and surface instabilities, with larger wavelengths resulting from a combination between the two effects.

\section{Conclusions}

We overtook the modelling of nuclear jets in heavy-ion collisions in the Fermi-energy range, i.e. the arising of collimated streams of low-density matter.
	At variance with other dissipative mechanisms in nuclear dynamics, the jet formation has not been explicitly addressed so far, even though it has been reported in experiments without being identified as a specific process of instability.
	While a simplified description restricted to the one-body contribution results insufficient, we found that nuclear jets
 are described when collisional correlations and fluctuations in full phase space are included, relying on a full solution of the Boltzmann-Langevin equation.
	In this case, we evidence for the first time the concurrent role of volume and shape instabilities. Our calculations show that the disassembling of the jet occurs mainly by a combination of volume instabilities over a range of densities which extends below $1/4$ of saturation density, rather than by Rayleigh surface instabilities. 
	The mechanism favours the production of small fragments, down to light clusters.
	Only a residual surface contribution affects the largest wavelengths, indicating a possible transition from volume to surface instabilities.
	The vanishing contribution of cohesive forces in the clusterisation of the jet, inspires a suggestive analogy between nuclear jets and granular streams of dry sand.
	In new forthcoming experiments where isotopic identification and particle correlations will be measured, also non-equilibrium features like the fragment growth time investigated in this work could be more directly accessed, opening novel frontiers.

\bigskip
{\flushleft\bf Acknowledgments}
\bigskip

Research was conducted in the scope of the International Associated Laboratory (LIA) COLL-AGAIN.

\appendix
\section{Some numerical details \label{supplemental}}

The Boltzmann-Langevin (BL) treatment of Eq.~(\ref{eq:BL}) as well as applications to instabilities and fluctuating behaviours are described in details in ref.~\cite{Napolitani2017}.
Through a Wigner transform, some correspondence can be established with the stochastic TDHF~\cite{Ayik1988,Lacombe2016} scheme.
	 The reduced one-body phase-space density $f^{n}$ in Eq.~(\ref{eq:BL}) replaces in fact the Slater representation in the TDHF approach and corresponds to the Fermi statistics at equilibrium.
The residual contributions are replaced by modified Uehling-Uhlenbeck (UU) terms, where each single in-medium collision event acts on extended equal-isospin phase-space portions, large enough so that the occupancy variance in $h^3$ cells corresponds to the one associated with the scattering of two nucleons; this variance should equal $f(1-f)$ at equilibrium, in order to strictly avoid any violation of Pauli blocking at each single scattering event~\cite{Rizzo2008}.
	A solution of the BL equation was obtained in full phase space, resulting in the following set of BLOB equations~\cite{Napolitani2013,Napolitani2017}:
\begin{eqnarray}
	&&\frac{\partial f^{(n)}}{\partial t} - \{h^{(n)} , f^{(n)}\} 
		= I_{\textrm{UU}}^{(n)} + \delta I_{\textrm{UU}}^{(n)} = 
\notag\\
	&&= g\int\frac{\diff\vecp_b}{h^3}\,
	\int
	W({\scriptstyle\A\B\leftrightarrow\C\D})\;
	F({\scriptstyle\A\B\rightarrow\C\D})\;
	\diff\Omega
\;,
\label{eq:BLOB}
\end{eqnarray}
where $g$ is the degeneracy factor, $W$ is the transition rate in terms of relative velocity between the two colliding phase-space portions, and $F$ handles the Pauli blocking of initial and final states over their full phase-space extensions.

	All simulations presented in this study use a simplified SKM$^{*}$ effective interaction~\cite{Guarnera1996,Baran2005} with compressibility modulus $k=200$MeV and a linear parameterisations for the surface symmetry energy. 
$\Ntest=40$ test particles per nucleons are used to sample the mean field.
	A screened in-medium n-n cross section (from ref.~\cite{Danielewicz2011}) is used.
	The calculation is carried on till no new fragments appear and no further than 280 fm/c; it is then rewinded back to the time of the last fragment separation $t_{\textrm{stop}}$, which is different for each event.

	The density dependencies can be included in the surface tension $\gamma$ in Eq.~(\ref{eq:Rayleigh}) as suggested in refs.~\cite{Iida2004,Horiuchi2017}, so that
%
\begin{equation}
	\frac{\gamma(\rho,\beta,T)}{\gamma_{\textrm{sat}}} 
	\approx 
	F_T	
	\Big[ 1 -c_{\textrm{sym}}\beta^2 -\chi\Big(1-\frac{\rho}{\rho_{\textrm{sat}}}\Big) \Big]
	\;,\label{eq:surfacetension}
\end{equation}
where, according to the SKM$^{*}$ interaction, 
$\gamma_{\textrm{sat}}\!=\!\gamma(\rho_{\textrm{sat}},\beta\!=\!0,T\!=\!0)\approx 1 \text{MeV\,fm}^{-2}$,
$\rho_{\textrm{sat}}\approx 0.16 \text{fm}^{-3}$,
$c_{\textrm{sym}}\approx1.9$
and 
$\chi=(\rho_{\textrm{sat}}/\gamma_{\textrm{sat}})\partial_\rho \gamma|_{\rho_n=\rho_p=\rho_{\textrm{sat}}/2}\approx1.16$.
	$F_T$ further introduces a temperature correction which, according to the proposal of ref.~\cite{Ravenhall1983}, imposes that the surface tension should vanish at the critical temperature of the nuclear liquid-gas phase transition ($T_{\textrm{c}}\sim$18 to 20 MeV).
	In the conditions of the present study, corresponding to a temperature of about 3~MeV, the correction is still negligeable, so that $F_T\approx1$.
	Diffuseness and viscosity~\cite{Baldo2012}, which have antagonist contributions~\cite{Brosa1990}, as well as geometric distortion are neglected. 

	In Eq.~(\ref{eq:spinodal}), the ultraviolet cutoff is determined by the range of the nuclear interaction through a term related to surface tension and diffusivity. 
It consists of a Gaussian smearing $g(k)$ with a width of around 0.8 to 0.9fm~\cite{Napolitani2017}.


%
%
%
%

\end{document}